\documentclass{article}
\usepackage{fullpage}
\usepackage{url}
\usepackage{graphicx}
\usepackage[draft,inline,nomargin,index]{fixme} 
  \fxsetup{theme=colorsig,mode=multiuser,inlineface=\sffamily,envface=\itshape} 
  \FXRegisterAuthor{sv}{asv}{Suresh} 
  \FXRegisterAuthor{jp}{ajp}{Jeff} 

\newcounter{Ccount}
\newcommand\chal[1]{\vspace{.05in}\noindent {\sffamily \textbf{Cyclops \addtocounter{Ccount}{1}\arabic{Ccount}:} #1}\vspace{.05in} }

\title{Rethinking Abstractions for Big Data: \\ Why, Where, How, and What}
\author{ \small
Mary Hall,
Robert M. Kirby,
Feifei Li, 
Miriah Meyer,
Valerio Pascucci,
Jeff M. Phillips, 
\\ \small
Rob Ricci, 
Jacobus Van der Merwe,
Suresh Venkatasubramanian,  
\\ \small University of Utah
} 

\begin{document}
\maketitle


Big data refers to large and complex data sets that, under existing approaches, exceed the capacity and capability of current compute platforms, systems software, analytical tools and human understanding~\cite{Mckinsey}. 
Numerous lessons on the scalability of big data can already be found in asymptotic analysis of algorithms and from the high-performance computing (HPC) and applications communities.  
However, scale is only one aspect of current big data trends; fundamentally, current and emerging problems in big data are a result of unprecedented \emph{complexity}---in the structure of the data and how to analyze it, in dealing with unreliability and redundancy,
in addressing the human factors of comprehending complex data sets, in formulating meaningful analyses, and in managing the dense, power-hungry data centers that 
house big data.

The computer science solution to complexity is finding the right abstractions, those that hide as much triviality as possible while revealing the essence of the problem that is being addressed. 
The ``big data challenge'' has disrupted computer science by stressing to the 
very limits the familiar abstractions which define the relevant subfields in data analysis, data management and the underlying parallel systems.
Efficient processing of big data has shifted systems towards increasingly heterogeneous and specialized units, with resilience and energy becoming important considerations. 
The design and analysis of algorithms must now incorporate emerging costs in communicating data driven by IO costs, distributed data, and the growing energy cost of these operations.  
Data analysis representations as structural patterns and visualizations surpass human visual bandwidth, structures studied at small scale are rare at large scale, and large-scale high-dimensional phenomena cannot be reproduced at small scale. 

As a result, not enough of these challenges are revealed by isolating abstractions in a traditional software stack or standard algorithmic and analytical techniques, and attempts to address complexity either oversimplify or require low-level management of details.
The authors believe that the abstractions for big data need to be rethought, and this reorganization needs to evolve and be sustained through continued cross-disciplinary collaboration.

In what follows, we first consider the question of \emph{why} big data and why now.  We then describe the \emph{where} (big data systems), the \emph{how} (big data algorithms), and the \emph{what} (big data analytics) challenges that we believe are central and must be addressed as the research community develops these new abstractions. 
We equate the biggest challenges that span these areas of big data with big mythological creatures, namely cyclops, that should be conquered.  

\subsection*{Why Big Data, And Why Now?}

We argue that big data is not simply the familiar outcome of constant improvements to processing, storage, and networking---it is something qualitatively different with a novel set of problems that are the product of three factors:
(1) The \emph{availability} of large data sets---not only to communities that
are accustomed to dealing with them, but also to commercial and academic groups
that have historically been unable to collect or store large amounts of data.
The consequence is that what constitutes ``big'' varies greatly among 
communities---the common theme is that the data is large enough to place
a strain on the existing techniques used to manage, process, and interpret it.
(2) The \emph{capacity} to process and store large data sets---a result not only
of predictable advances in hardware technology, but also of disruptive changes 
to the models of how this capacity is made available. The ability to rent or
borrow computing and storage capacity at a variety of timescales changes the
basic economics of the value of data---data becomes more valuable when this
capacity is available on demand and at the exact scale required for the problem.
(3) The realization that there is \emph{economic or social benefit} to be
gained from these data sets, whether that be for the advancement of technology,
science, commerce, or the public good.
All of these factors already exist in one form or another; what is new is that advances in each of these areas have converged.  The result is a much broader cross-disciplinary interest that extends across and beyond the communities that have traditionally looked at problems
related to data.

\begin{figure}
\begin{center}
\includegraphics{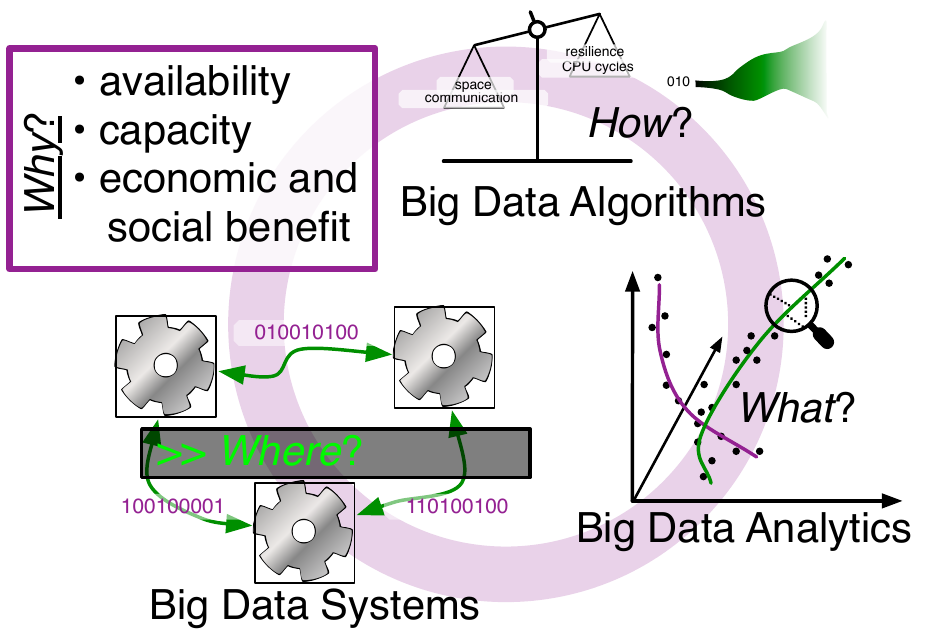}
\end{center}
\caption{\label{fig:WWHW} The why, where, how, and what of big data.}
\end{figure}

\section{Big Data Systems}
\label{sec:big-data-systems}
Big data places new strains on the compute platforms \emph{where} the analysis
will be performed:  
the layers of architecture, OS, compiler, and programming languages which allow programmers to take advantage of a variety of underlying platforms.
The high-performance computing (HPC) community has historically been the mainstay of scalable compute systems, along with closely related work in grid computing~\cite{Globus}.  More recently, a second community, led by companies such as Google, Yahoo!, and Amazon have been building a new set of systems and abstractions for big data processing.  MapReduce~\cite{MapReduce} is the most notable example of this latter trend.
These two communities have developed what may appear to be distinct and contrasting perspectives on extreme-scale systems, but both approaches have clear strengths.  Perhaps surprisingly, these domains are converging towards common concerns, and we believe that a mixing of their abstractions brings both challenges and opportunities.

\subsection*{Performance vs. Productivity} 
A key area where divergence exists today is in programming models.
The HPC community uses fairly low-level programming models that provide control of hardware features such as message-passing across distributed memories and explicit threading to control processor cores, designed for expert users who demand high performance
over programmability.  In contrast, MapReduce~\cite{MapReduce} is a commonly-used large-scale data analysis programming model, where programmers work with two simple abstractions: ``mapping'' computation across individual key-value items, and ``reducing'' the intermediate results to a final output or another intermediate state.
The simplicity of this programming model, which favors programmer productivity over performance, has resulted in rapid and widespread adoption.

\begin{center}
\emph{While today's systems make different choices in the spectrum between performance and productivity,  the tradeoffs these communities are facing in the future are bringing them closer together.}
\end{center} 

The HPC community, in its march to exascale---the ability to process a quintillion operations per second---is exploring more productive programming environments that hide the increased complexity of new energy-efficient hardware features such as heterogeneous processing logic, non-uniform latency memory structures, configurability, and billion-way parallelism.
The data analysis community finds the MapReduce paradigm too restrictive for some computations, and is exploring new programming models to facilitate other data analytics solutions, e.g., Dremel~\cite{Dremel} (for queries), GraphLab~\cite{GraphLab} (for machine learning and optimization), and Giraph (for graph processing).
Performance and efficiency will also become more important for extreme-scale data analysis, particularly in data centers. 
Thus, the need to ``dial-in'' to appropriate levels of performance, productivity, and abstraction
exists in both communities.
\emph{Multi-resolution programming models} have been proposed to permit different users with varying programming expertise to use the same programming system at different levels of abstraction; experts can have control of application mapping to hardware, while more na\"{i}ve users can lean more heavily on automation and general-purpose solutions~\cite{DOE-exaprog}.  A common example of such a multiresolution system relies on a \emph{domain-specific} programming language or library that specializes the expression and mapping of particular domains so that more na\"{i}ve users can capitalize on efforts of expert users to manage system complexity.  
Meanwhile Hadoop~\cite{YARN} (the open source version of MapReduce and its relatives) now permits some version of message passing to allow advanced users to further optimize some key operations outside the basic, but restricted, paradigms.  
Thus, while the applications and programming solutions may vary significantly, the current and future experiences of successfully deploying domain-specific systems should provide general lessons towards addressing these challenges in both the HPC and data analysis communities.  

\subsection*{Toward Resilience and Energy-Efficiency}
The application programming environment is only one part of the overall computing environment; it is layered on the run-time system, operating system, file system and networking layers. While today, the choices and abstractions made for these layers appear to differ between the HPC and data analytics communities, again we see initial steps towards convergence. 
Historically, 
the HPC community has used expensive hardware with low failure rates, and relies on heavyweight checkpoint and restart of long-running applications to tolerate infrequent errors.  In contrast, data centers are designed to expect frequent failures; a typical strategy is to use
low-cost hardware components that have poor failure rates, and to build flexible, distributed failure recovery schemes as part of the systems or application layers.
With supercomputers growing dramatically in component counts and checkpoint-restart becoming too expensive relative to compute time,  the HPC community is by necessity moving towards a more resilient approach that is tolerant or able to recover from errors~\cite{exasys,DOE-exaprog}. 
Similarly, for both supercomputing facilities and data centers, power consumption will dominate the operational costs, making energy efficiency a first-class concern for both communities.
Rather than relying completely on hardware solutions for energy efficiency, more aggressive management of data movement (which dominates energy costs~\cite{exasys}), must be addressed at various levels of the software stack.
The clear conclusion is that, regardless of the specific application-level abstractions used for computation on big data, there will be common needs for both resiliency and power efficiency; this points to the need to re-architect systems software to addresses these challenges.

\section{Big Data Algorithms}

The rise of big data has disrupted the algorithmic paradigms that model \emph{how} data is represented and processed.  
Fundamentally, algorithm design is about solving concrete problems within the confines of an abstract framework.  This framework should elegantly capture the key trade-offs between resources, whether they be time, space, precision, random bits, or communication, but hide the implementation details from the algorithmic designer.  For years, computer systems have managed to handle these details behind the scenes (from an algorithmist's perspective), and keep the balance in check.  
However, the evolution of the true trade-offs has reached a breaking point.  
Most evidently, new computational models are now needed, ones that may deal with an ever-growing list of competing resource types.  
But also, the infallibility of input data's accuracy, which had been masked by wise choices of concrete problems, is becoming a first-order concern.  
Both of these issues are forcing changes in design principles, and for algorithm designers' efforts to be most effective, new abstractions that are at the same time simpler and richer are required.

\subsection*{New Models for Processing Large Data}

For nearly 70 years, algorithms have been developed under variants of the Von
Neumann model~\cite{VonNeumann}. The key feature of this model is the assumption that accessing data and performing an operation on it took the same time. While this
assumption has never  held on any actual architecture, cache hierarchies and fast memory were able to make this an accurate model of reality \ldots until now.

Increasing data sizes and the dramatic increase in compute speeds have changed the balance between computation and access. Processing data is now far cheaper than accessing it.  As a consequence, data access has itself become either an expensive resource to be optimized (as seen in the very successful external memory and cache-oblivious models of computation) or one that is extremely constrained, as in the \emph{streaming model}~\cite{streaming} in which algorithms are only permitted to make a single pass over the data, and can only store a tiny fraction of what they read. 

While these models have been quite effective at modeling the problems of data access on a \emph{single processor}, it has been much harder to adapt them to the challenges of dealing with multiple processors on vast amounts of distributed data. Whether we are programming
a GPU, a multicore processor, a cluster of thousands of processing nodes, or even computations across distributed data servers, we have to model \emph{communication} (between compute nodes, or between the memory hierarchy and nodes) rather than \emph{access}. 

\begin{center}
\emph{Modeling the interplay between communication and computation is perhaps the biggest challenge for algorithm design in the modern era. }
\end{center}

Valiant points out~\cite{Bridging} that an ideal computational model ``bridges in a performance-faithful manner what the hardware executes and what is in the mind of the software writer.''   Where do we stand with respect to this ideal?   
As we discussed in Section~\ref{sec:big-data-systems}, the tradeoff between performance
and productivity reflects the disconnect between the ``truth'' of the hardware and the various programming abstractions used to interface with it. 
This disconnect transfers to theoretical models as well. While there are now numerous models for thinking abstractly about big data systems, they are inspired by specific programming interfaces in use rather than by a deeper understanding of underlying hardware and software realities. 
One notable exception is Valiant's own recent work on bridging models for multicore systems. But even this work is limited to a homogeneous hierarchy of (memory) layers and does not capture the full complexity of modern large-scale distributed parallel systems. 

We should note here that there are many other factors that play an important role in the design and modeling of large-data systems. In Section~\ref{sec:big-data-systems} for example, we discuss heterogeneity, resilience and energy-efficiency, among others. While we believe that communication is the most natural \emph{resource} for modeling, we expect that algorithmic principles that address the above concerns will continue to be of interest. 
Indeed, as we have seen with the very successful streaming model, the key challenge in designing an effective theoretical framework will not be exhaustiveness, but a focus on the ``right'' resources to optimize.

\subsection*{Algorithmically Managing Inaccuracies}

Big data is often noisy.  Actually, data has always been noisy, but the problem is more apparent with large data since we often observe the same object multiple times, and these observations can have conflicting values.  Furthermore, the data has reached a size where it is possible to effectively model this uncertainty.  This realization has two consequences.  First, precision of solutions beyond the error tolerance of the input data is meaningless; approximate solutions can be used in place, as long as they have guarantees within this tolerance.  
Second, this input data uncertainty should be analyzed, in particular with regard to its affect on the output of a given task.  
That is, the input, the intermediate structure, and the output of these noisy data sets should have rich representations either describing the distribution of possible solutions, or tolerances of the worst case values.  

However, computing on such complex representations, especially when the data is big, requires new ideas.  
One approach is to \emph{squash} the data to a convenient smaller representation that also captures this data uncertainty.  In fact, this process will often introduce further inaccuracies, but these can often be carefully bounded and modeled.  
This approach yields two challenges: the first is efficiently computing this complex but concise representation from an enormous data set, and second is maintaining such a rich representation as this data is processed through several analysis steps.  With each step, we would ideally like to rely on an existing algorithmic technique, but not have the inaccuracies increase and thus propagate at a larger level.  

Currently, there are Monte Carlo approaches which are quite general, as well as drastically more efficiently techniques (often growing out of streaming algorithms~\cite{streaming}) but that typically apply to only specific scenarios.    
A generally and efficient algorithmic framework that can be abstracted to many types of complex data is a important challenge.

\section{Big Data Analytics}

Data science, the emerging scientific paradigm of discovering hypotheses from large data corpora, has turned the natural progression of science on its head. Instead of painstakingly designing hypotheses and testing them, it is now possible to generate hypotheses automatically by sifting through giant data sets.

The danger of this approach is the phenomenon of \emph{multiple testing} (or more colorfully, the green jelly bean problem \url{http://xkcd.com/882}), where if enough hypothesis are considered separately, eventually one observed effect may look statistically significant without being true.  
This problem is all the more serious with large and complex data because the algorithms that generate these hypotheses can be opaque, and the data itself can overwhelm our ability to process and visualize it. Moreover, the number of features in the data can overwhelm most
procedures designed to analyze them.  

The challenge of big data analytics therefore is to determine \emph{what} information and structure really lies in these large, feature-rich data sets, and which models that can be evaluated efficiently and accurately, and visualized to provide confirmation of the learned phenomena.

\subsection*{Structure that is \emph{not} Everywhere}

The curse of dimensionality refers to the exponential increase in complexity of algorithms as the number of features (the dimensions) of data increases.  But this phenomenon is not just algorithmic; in high dimensions the meaningful patterns in data become harder to distinguish from random artifacts, regardless of the efficiency of the algorithm used.  
However, most natural processes that generate data tend to be less complex, and this has spurred the development of methods that \emph{assume} the data lives in a low-dimensional subspace, and find patterns conditioned on this or other assumptions.  While it can be hard to verify that the data satisfies these constraints, regularization methods can help nudge algorithms to look for such structures, and in some cases cross-validation can confirm the validity of found structures.  

Another path for rigorous analysis of complex data is through the study of which summaries with worst-case guarantees can be attained, and understanding the trade-off between error tolerance and size.  Limits on these summaries can imply that even big data sets can not provide more than a fixed error tolerance for certain properties, and limit overzealous modeling and over-fitting.  In simulation data, modeling error can be much more dramatic, since data sets are the output of the models.  This requires efficient and early detection of structural anomalies to short-circuit the expense of regenerating the data, and efficiently maintainable summaries can be the key to this.  

As data becomes more complex, the models must adapt as well. Over the years we have seen models for measuring data evolve from simple linear spaces to infinite dimensional function spaces. We are now seeing a further evolution into \emph{multi-scale} representations of
data, where ``low-dimensional spaces'' combine ``low dimensionally'' to create complex structure.  But it still remains a challenge to automatically adapt this model complexity appropriately to the scale and structure of the data.

\subsection*{Learning Complex Structures}
\label{sec:machine-learning}

Advanced data models represent one component of the challenge of dealing with high-volume feature-rich data.  In addition, the model learning strategies themselves are being revisited in the face of the scale and complexity of data. 
A learning task typically proceeds by training a model and then testing it (in batch mode) or building a model that is updated as new data appears (in an online setting). The training  involves nontrivial optimization and often relies on labeled data and a selected set of features on which to build a model. 
But managing the choice of features and labeling decisions among large, complex, and distributed data requires new insights beyond the related algorithmic challenges.  

Without a global view of the data, ensemble methods will become very important; these methods combine several models, perhaps learned on different views of the data, into a single global model.  These have already been successful in large-scale learning applications, such as in the Netflix challenge.  Another approach is through transfer learning where model parameters (rather than data) are transferred between distributed entities.  For instance, the GraphLab system~\cite{GraphLab}  has (among other things) implemented parallel versions of belief propagation in a way that generalizes MapReduce.  Beyond these approaches, active learning and multi-task learning make use of auxiliary data sources to minimize the cost of acquiring and using labelled data, and a challenge in ongoing work is to adapt them to large distributed data sets.

\subsection*{Information Bandwidth Overload}

When analyzing big data, classic two-dimensional statistical plots are often insufficient for exploring and understanding the complex patterns and relationship embedded within. Furthermore, deciphering raw data and computational results through visual representations is too often done as the final step in a complex research process, with tools that are rarely specific to the task. 

\begin{center}
\emph{To truly close the loop in data science, human-understandable representations of data must be made available throughout the analysis pipeline to help guide analysts in making decisions and discoveries.}
\end{center} 
Interactive visualizations support this process by allowing vast amounts of information to be encoded, and rely on our powerful perceptual systems to pull out interesting trends and structures.  Interactive, flexible, and sophisticated visualization tools allow analysts to validate data and models, to derive new hypotheses, and to make important discoveries. 

Interactive visualizations have started to replace the classic static images printed on paper, giving the viewer the ability to navigate multiple views of a data set. By linking together these different views through user interactions, new paradigms of data exploration are possible; for example, a large, multidimensional data set can be visualized by integrating a high-level summary view with more detailed, fine-grained views of subsets of the data. This allows deeper, more specific exploration of the data without compromising its breadth. This design pattern, called \emph{overview+detail}, is found in many visualization tools for big data.  Abstractions like this are vital for designing effective visualization tools, but they need to be specialized within the heterogeneous landscape of applications and data, need to have safeguards warning of multiple testing problems, and need to be built on top of efficient indexing structures to allow for useable levels of manipulation.

Effective visualization tools need to carefully straddle the competing demands of generality and specificity. On one hand, we need to devote time and resources to create tools that will support many different analysts in many different application areas---we need to find broadly applicable visualization abstractions. On the other hand, however, individual questions and inquiries often require specialized data and visual abstractions to tackle a specific problem, and individual cognitive differences between analysts can effect the interpretation of a visualization. Understanding when, where, and how visualizations can impact analysis of big data are ongoing research questions that draw on knowledge from computer science, cognitive science, and design.

\section*{Challenges}

As a take-away message, we summarize the big challenges outlined in this document on resolving the influx of complex big data with new abstractions.  To be clear, there are many other challenges associated with big data, especially those dealing with the social, legal, and economic aspects. 

The cyclops were \emph{big} one-eyed creatures from Greek mythology and needed to be conquered by ``heroes'' such as Odysseus.  Thus we identify the big challenges of big data  with cyclops: 
\emph{Brontes} the ``thunderer," \emph{Steropes} the ``flasher," \emph{Polyphemus} the ``shepherd," and \emph{Arges} the ``brightener."

\chal{[Brontes: System Abstractions] Design abstractions and languages for big data systems that ``slide'' gracefully between exposing a simple programming model that results in adequate performance for a broad class of programmers, yet making available lower-level details when necessary to maximize performance.} 

\chal{[Steropes: Algorithmic Models] Converge to computational abstractions which closely translate to and between the various evolving big data systems, and capture whichever emerging costs (e.g., communication, power, resiliency, precision, heterogeneity) dominates this new landscape.}  

\chal{[Polyphemus: Uncertainty Management] Assess the uncertainty and confidence in big data corpuses, and develop a framework for efficiently managing, processing, representing, and visualizing its effect up to the resolution at which it is reliable.}

\chal{[Arges: Reliable Structure]  Efficiently identify inherent low-dimensional or core structure from complex data without over-fitting, and represent this structure so it can be easily verified, analyzed, and visualized at multiple scales.}


Although these problems have been separated into categories, clearly the way forward should be a joint effort along all of these fronts.  Breakthroughs or resolutions in one area will have tremendous influence in others.  So it is paramount that system designers, algorithm experts, and data analysts work closely together to bring forth a new and exciting era of big data computing.  

\bibliographystyle{plain}
\bibliography{bgdata}

\end{document}